\documentclass[a4paper, aps, prb, amsmath, amssymb, superscriptaddress, preprintnumbers, onecolumn, 12pt]{revtex4-1}
\usepackage{graphicx,xcolor}
\usepackage{epstopdf}
\usepackage{mathptmx, textcomp}
\usepackage[utf8]{inputenc}
\usepackage[T1]{fontenc}
\usepackage{braket}
\usepackage{amsmath}
\usepackage{changes}
\usepackage{upgreek}
\usepackage{color}
\usepackage{xcolor} 
\usepackage[UKenglish]{babel}
\usepackage{array}

\definechangesauthor[name=Daniel,color=blue]{dh}
\definechangesauthor[name=Daniel,color=violet]{dh2}
\definechangesauthor[name=JHD,color=red]{JHD}
\DeclareMathAlphabet{\mathcal}{OMS}{cmsy}{m}{n}

\DeclareSymbolFont{alphabets}{\encodingdefault}{cmr}{m}{it}

\SetSymbolFont{alphabets}{normal}{\encodingdefault}{cmr}{m}{it}

\SetSymbolFont{alphabets}{bold}{\encodingdefault}{cmr}{b}{it}

\DeclareMathSymbol{v}{\mathalpha}{alphabets}{"76}

\begin{document}
	
\title{Reaction kinetics of ultracold molecule-molecule collisions}

\author{Daniel K. Hoffmann}
\affiliation{Institut f\"{u}r Quantenmaterie and Center for Integrated Quantum Science
and Technology (IQST), Universit\"{a}t Ulm, D-89069 Ulm, Germany}
\author{Thomas Paintner}
\affiliation{Institut f\"{u}r Quantenmaterie and Center for Integrated Quantum Science
and Technology (IQST), Universit\"{a}t Ulm, D-89069 Ulm, Germany}
\author{Wolfgang Limmer}
\affiliation{Institut f\"{u}r Quantenmaterie and Center for Integrated Quantum Science
and Technology (IQST), Universit\"{a}t Ulm, D-89069 Ulm, Germany}
\author{Dmitry S. Petrov}
\affiliation{LPTMS, CNRS, Univ. Paris Sud, Universit\'{e} Paris-Saclay, 91405 Orsay, France}
\author{Johannes Hecker Denschlag}
\affiliation{Institut f\"{u}r Quantenmaterie and Center for Integrated Quantum Science
and Technology (IQST), Universit\"{a}t Ulm, D-89069 Ulm, Germany}

\begin{abstract}
{\textbf{
Studying chemical reactions on a state-to-state level  tests and improves our  fundamental understanding of chemical processes.
For such investigations it is convenient to make use of ultracold  atomic and molecular reactants as they can be prepared in well defined internal and external quantum states
\cite{Kre08,Car09,Que12,Ric15}.  In general, even cold reactions have many possible final product states \cite{Zah06,Sta06,Fer08,Muk04,Sya06,Zir08,Osp10,Ni10,Mir11,Wan13,Dre17} and reaction channels are therefore hard to track individually \cite{Wol17}.
 In special cases, however, only a single reaction channel is essentially participating, as observed e.g. in the  recombination
of two atoms forming a Feshbach molecule \cite{Joc03,Cub03,Reg04} or in atom-Feshbach molecule exchange reactions \cite{Rui17,Kno10}.
Here, we investigate a single-channel reaction of two Li$_2$-Feshbach molecules where one of the molecules dissociates
into two atoms
$2\mathrm{AB}\Rightarrow \mathrm{AB}+\mathrm{A}+\mathrm{B}$. The process is a prototype for a class of four-body collisions where two reactants produce three product particles.
We measure the collisional dissociation rate constant of this process as a function of collision energy/ temperature and scattering length. We confirm an Arrhenius-law dependence on the collision energy, an $a^4$ power-law dependence on the scattering length $a$ and determine a universal four body reaction constant.}}
%\replaced[id=dh]{From the scattering length scaling we estimate the universal scaling of the three-body association rate constant in a collision of three distinguishable particles.}{}.
\end{abstract}
\maketitle
We carry out the experiments with an ultracold, mixed gas of  Li$_2$ Feshbach molecules and unbound Li atoms for which the reaction dynamics is governed by a detailed balance. 
By measuring the particle numbers for different settings in and out of equilibrium we determine the reaction rate constants for collisional dissociation and association of a  Li$_2$ molecule. 
We find that the dissociation rate constant strongly depends on the  temperature and the scattering length, in agreement with theoretical predictions.

The initial atomic and molecular sample is prepared from an ultracold gas of $N_{tot}=2.6\times10^{5}$ fermionic $^6$Li atoms which consists of a balanced mixture of the two lowest hyperfine states $|m_F=\pm1/2>$ of the electronic ground state.
In the vicinity of the Feshbach resonance at $B_0=832.2\,\mathrm{G}$ (see ref.\cite{Zur13}) exothermic three-body recombination can convert pairs of $|-1/2>, |+1/2>$ atoms into weakly-bound Feshbach molecules, each of which is in the same internal quantum state.
The process is reversible and a Feshbach molecule can dissociate again into the unbound $|-1/2>, |+1/2>$ atoms via an inelastic, endothermic collision with another molecule or atom.
At thermal equilibrium  balance of the back and forth reactions is established.
This balance is a function of the particle densities, temperature, molecular binding energy, and scattering length, all of which can be controlled in our setup via confinement, evaporative cooling, and by choosing a magnetic offset field $B<B_0$.
Our trap is a combination of a magnetic trap and an optical dipole trap and is cigar shaped.
The trap has a depth of $U_0=21\,\mathrm{\upmu K}\times k_B$, corresponding to a radial (axial) trapping frequency of $\omega_r = 2\pi\times0.99\mathrm{kHz}$ ($\omega_{ax}=2\pi\times21\mathrm{Hz}$), respectively.
We use evaporative cooling to set the temperature to approximately $1.2$ to $1.3\,\mathrm{\upmu K}$.
At this temperature  80$\%$ to 90$\%$ of all atoms are bound in Feshbach molecules within the $B$-field range of $705$\,G to $723\,\mathrm{G}$ of our experiments,  corresponding to a binding energy $E_b$ between $6$ and $10\,\mathrm{\upmu K}\times k_B$ (see Methods).	
We note that at these settings where $T \ge T_F$ ($T_F$ is the Fermi temperature) and $E_b>k_BT_F$, quantum degeneracy only plays a negligible role for the reaction kinetics.

 \begin{figure}[tbp]
 \centering
 \includegraphics[width=1\textwidth]{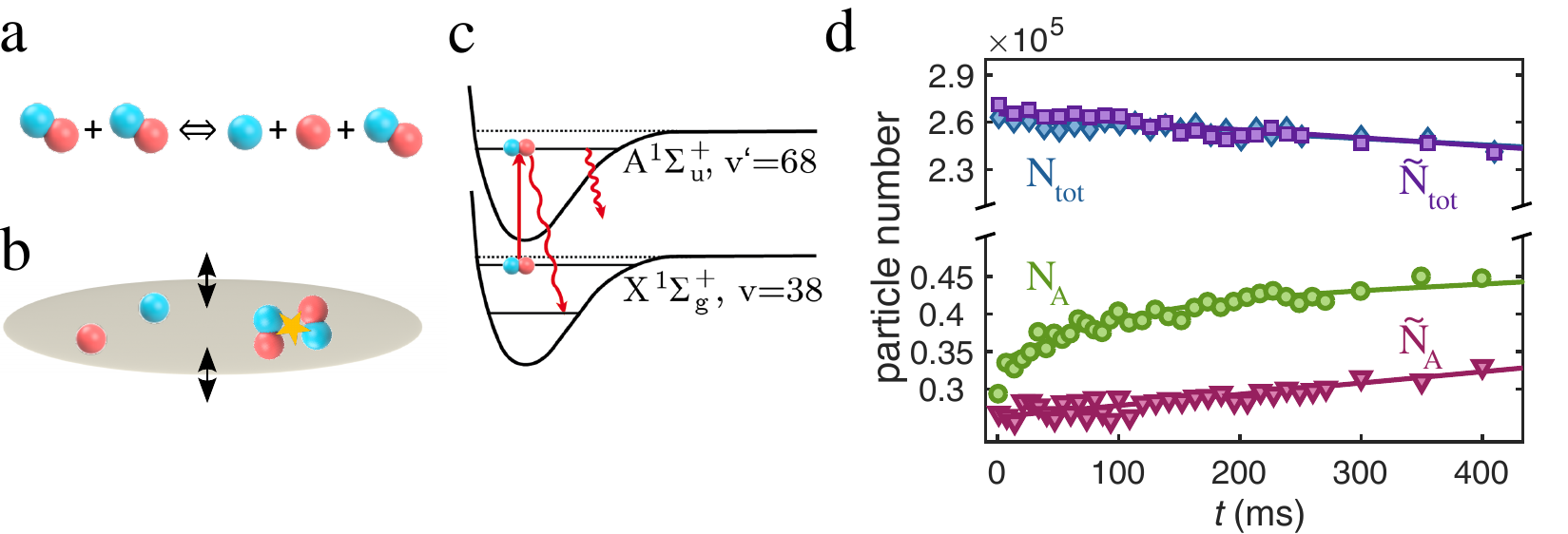}
 \caption{\textbf{Molecule dissociation dynamics. a,}
 Detailed balance of collisional dissociation and association of dimers.
 \textbf{b,}  A parametric heating pulse triggers the reaction dynamics. \textbf{c,} As part of the detection scheme, the Feshbach molecules which have a large admixture of the $X^1\Sigma_g^+,v=38$ state are optically pumped to undetected atomic or molecular states via the intermediate level $A^1\Sigma_u^+,v'=68$.
   \textbf{d,} Measurement of dissociation dynamics at $709\,\mathrm{G}$.
   Lower part: Circle (triangle) symbols show the number of unbound atoms $N_{A}$ ($\tilde{N}_{A}$) for variable holding time $t$ with (without) initial parametric heating pulse.
    Upper part: Diamond (square) symbols show the total particle number $N_{tot}$ ($\tilde{N}_{tot}$) with (without) parametric heating pulse. }
 \label{fig:DissoResults}
 \end{figure}

In our first experiment we suddenly raise the temperature of the gas
%We start by performing an out-of-equilibrium  experiment
%by suddenly raising the temperature of the gas
 using an excitation pulse of parametric heating. This shifts the gas out of thermal equilibrium and the gas responds by collisionally dissociating a part of its molecules,   (see fig. \ref{fig:DissoResults}a). For this, we modulate the  dipole trap depth (see fig. \ref{fig:DissoResults}b) with frequency $\omega_{heat}\approx1.7\omega_r$ and amplitude $\Delta U=0.21U_0$ during a period $t_{p}=20\,\mathrm{ms}$.
After the excitation atoms and dimers thermalize on a time scale of a few milliseconds via elastic collisions, whereas the chemical equilibrium requires a much longer time of $150\,\mathrm{ms}$.

To investigate these dynamics  we measure how  the number of molecules $N_M$ and the number of unbound atoms $N_{A}$ change as a function of time.
We measure $N_{A}$ by using standard absorption imaging.
However, prior to the imaging  we first %\replaced[id=dh]{}{we}
remove all Feshbach molecules from the gas. %\replaced[id=dh]{}{to avoid spurious signals}.
For this, a resonant laser pulse transfers the molecules to an electronically excited molecular state A$^1\Sigma^+_u, v' = 68$ which subsequently decays with in a few ns to undetected atomic or molecular states \cite{Pat05,Pai17} (see fig. \ref{fig:DissoResults}c), see Methods.
The laser pulse has a duration of $0.5\,\mathrm{ms}$ which is short compared to the reaction dynamics.
In order to determine $N_M$ we measure in a second run the total number of atoms $N_{tot}=2N_M+N_A$, whether they are bound or unbound, and subtract $N_{A}$.
For this we use again absorption imaging.
The Feshbach molecules are so weakly bound that the imaging laser resonantly dissociates them quickly into two cold atoms which are subsequently detected via absorption imaging{\cite{Ket08}}.

Figure~\ref{fig:DissoResults}d shows the measurements of $N_{A}$ and $N_{tot}$ as a function of holding time after the heating pulse.
While the total number of atoms $N_{tot}$ is essentially constant apart from some slow background losses, the atom number $N_{A}$ exhibits a 30\% increase in about 100ms which is the dissociation response of Li$_2$ molecules to the thermal pulse.
%In addition to
Besides this, $N_{A}$ also exhibits a slow, steady increase which we attribute to a background heating of the gas, e.g. due to spontaneous photon scattering of the dipole trap light (see Supplementary Information).
As shown by  $\tilde{N}_A$  in fig.~\ref{fig:DissoResults}d) this background 
 heating is also present  in the absence of the initial heating pulse.
Similarly,  the slow decay of $N_{tot}$ is also present without 
 the heating pulse (see $\tilde{N}_{tot}$ in fig.~\ref{fig:DissoResults}d).
It can be completely explained by inelastic collisions between molecules as previously investigated in ref. \cite{Bou04}. 

In principle, collisional dissociation in our experiment can be driven either by atom-molecule collisions or by molecule-molecule collisions.
We only consider molecule-molecule dissociation since its rate is about two orders of magnitude larger in our experiments than for atom-molecule dissociation with its known rate constant of\cite{Chi04} $C_1\approx10^{-13}\mathrm{cm^3/s}$ and given the fact that 
the mean density of atoms is a factor of ten smaller than for the dimers. 
In a simple physical picture, the suppression of the atom-dimer dissociation is due to the Pauli principle acting on the outgoing channel, which involves two identical fermionic atoms\cite{Pet03,DIn05}.
In the molecule-molecule collisional dissociation, the molecules can either dissociate into four unbound atoms,  $2\mathrm{AB}\Rightarrow 2\mathrm{A}+2\mathrm{B}$, or into two unbound atoms, $2\mathrm{AB}\Rightarrow \mathrm{AB}+\mathrm{A}+\mathrm{B}$.
However, since in our experiments the molecular binding energy $E_b$ is typically by a factor of 5 larger than the  thermal energy $k_BT$, the dissociation into four atoms comes at an additional sizeable energy cost and is therefore comparatively  suppressed by an Arrhenius factor of $\mathrm{exp}(-E_b/k_BT)\approx7\times10^{-3}$, see also \cite{Chi04}.
Therefore, to first order, we only need to consider dissociation into two atoms.
The evolution of the density $n_A$ of unbound atoms is then given by the rate equation,
\begin{equation}
\dot{n}_A=2\,{C}_2\,n_M^2-{R}_2\,n_A^2\,n_M/2 %\replaced[]{}{+2\,C_D\,n_M}
\label{eq:RateCh2dens}
\end{equation}
Here, $n_M$ is the  molecule density and ${C}_2$ (${R}_2$) are the rate constants of molecule dissociation (association).
A spatial integration of eq. \ref{eq:RateCh2dens} gives the rate equation for the number of unbound atoms,
\begin{equation}
\dot{N}_{A}=(4\pi^{3/2})^{-1}\frac{C_2}{\sigma_r^2\sigma_{ax}}\,N_{M}^2-2^{-7}(2\pi^2)^{-3/2}\frac{R_2}{\sigma_r^4\sigma_{ax}^2}\,N_A^2\,N_{M}
\label{eq:RateCh2I}
\end{equation}
where we assume a Boltzmann distribution in a harmonic trap.
Here, $\sigma_{r(ax)} = \sqrt{k_BT/2m\omega_{r(ax)}^2}$ denote the radial (axial) cloud width
of the molecular gas and $m$ is the mass of $^6$Li.  
Furthermore, in eq. \ref{eq:RateCh2I} we have used the fact that the  cloud size for the unbound atoms is $\sigma_{r(ax),A} = \sqrt{2} \sigma_{r(ax)}$.
By fitting  eq. \ref{eq:RateCh2I} to the data of fig. \ref{fig:DissoResults}d we can determine the rate coefficients to be 
${C}_2=(2.0\pm0.6)\times10^{-12}\,\mathrm{cm^3/s}$ and
	${R}_2=(4.1\pm1.2)\times10^{-22}\,\mathrm{cm^{6}/s}$.
%\replaced[id=dh]{$C_2$ and $R_2$}{$C_2, R_2$ and $C_D$}.
For the fit we use the measured widths $\sigma_{r(ax)}$ which turn out to be fairly constant during the holding time $t$ (a more detailed discussion will be given below).

  \begin{figure}[tbp]
  \centering
  \includegraphics[width=1\textwidth]{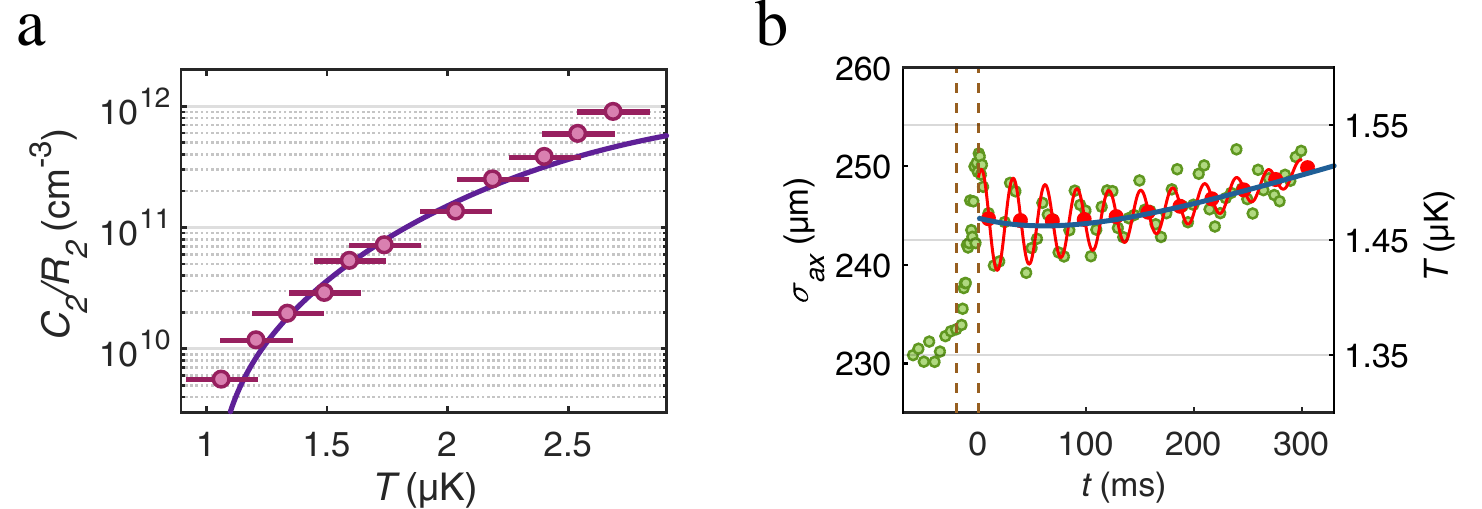}
  \caption{\textbf{Temperature dependence of the equilibrium state and temperature evolution.} \textbf{a,} The ratio $C_2/R_2$ (circles) is plotted as a function of temperature $T$ at $B=723\mathrm{G}$.
  The errorbars denote the $1\sigma$ uncertainty in the thermometry. The continuous line is a calculation without any free parameters (see text).
  \textbf{b,} Measured evolution of the axial cloud size $\sigma_{ax}$ (green circles) at $B=705\,\mathrm{G}$ after injecting a heat pulse during -20~ms$ < t < $ 0~ms (vertical dashed lines).
  The heat pulse abruptly increases the temperature $T$ and size $\sigma_{ax}\propto \sqrt{T}$. In addition it excites small collective breathing mode oscillations, see red line as a guide to the eye.
  The red dots mark the evolution of $\sigma_{ax}$ when averaged over one oscillation period. This evolution is well described by a model calculation (blue line) as described in the Supplementary Material. The temperature scale applies to the non-oscillatory part of the data. 
}
  \label{fig:TDepRates}
  \end{figure}

Next, we investigate how the reaction rates depend on temperature.
For this, it is convenient to study the atom molecule system in a state of near equilibrium, where $\dot{N}_A \approx 0$, i.e. $\dot{N}_A$ is much smaller
than the individual collisional dissociation/association rates of eq. \ref{eq:RateCh2I}. We can then set $\frac{{C}_2}{{R}_2}=2^{-5}(2\pi)^{-3/2}\frac{1}{\sigma_r^2\sigma_{ax}}\frac{N_{A}^2}{N_{M}}$. 
Thus, a change in the ratio ${C}_2/{R}_2$ due to a variation in temperature, can be experimentally observed in terms of a change of particles numbers and widths.
Figure~\ref{fig:TDepRates}a shows such measurements at $B=723\mathrm{G}$ ($\dot{N}_A$ was always at least a factor of ten smaller than the collisional dissociation/association rates).
Within the small temperature range between 1 and 3$\mu$K, the rate constant ratio $\frac{{C}_2}{{R}_2}$ increases by more than two orders of magnitude.
This result can be compared to a  prediction  based on statistical mechanics \cite{Chi04},
\begin{equation}
\frac{C_2}{R_2}  = h^{-3}(\pi m k_B T)^{3/2}e^{-E_b/k_B T}
\label{eq:ClTDep}
\end{equation}
which is shown in Fig. \ref{fig:TDepRates}a as a continuous line  with no adjustable parameters.
The agreement between experiment and theory is quite good. The strong increase of  $\frac{{C}_2}{{R}_2}$ with temperature is dominated by the Arrhenius law exponential $e^{-E_b/k_B T}$ which comes into play for the endothermic dissociation ($C_2$) but is absent for the exothermic recombination process ($R_2$).

The strong temperature dependence of the rate constants potentially has a strong influence on the reaction dynamics of our atom/ molecule system, as the chemical reactions change the temperature of the gas.
To quantify this influence, we take a closer look at the temperature evolution in our experiment by tracking the cloud size $\sigma_{ax} ~\propto \sqrt{T}$, see fig. \ref{fig:TDepRates}b.
Initially the system is in thermal equilibrium and the molecular cloud has an axial cloud size of about $\sigma_{ax} = 230\,\mathrm{\upmu m}$ which corresponds to a temperature of $T\approx1.3\,\mathrm{\upmu K}$.
The heating pulse, which starts at $t=-20\,\mathrm{ms}$ and ends at $t=0$, deposits thermal energy in the system.
Due to the fast elastic collisions of dimers and atoms the thermal energy deposition results in a fast increase of the cloud size of about $6\%$ which corresponds to a temperature increase of $\Delta T\approx0.15\,\mathrm{\upmu K}$.
In addition, the modulation of the dipole trap during the heating pulse  excites unwanted breathing mode oscillations in the cloud with a small amplitude of two percent.
The mean cloud size which is obtained by averaging over one oscillation (red circles in fig. \ref{fig:TDepRates}b) is almost constant within the first $150\,\mathrm{ms}$ after the heating pulse.
This might be at first surprising since one might expect the endothermic dissociation to considerably lower the temperature again. However, since the initial atom number is quite small, only a small amount of  molecules need to break up  to significantly increase the recombination rate $\propto N_{A}^2$ and thus to reach a new balance. Therefore only a small amount of the injected heat is consumed for the dissociation, corresponding to a small amount of cooling.
Moreover, this residual cooling is almost canceled by the background heating.
As a consequence the remaining temperature variation is less than 1\%.
For later times, $t>150\,\mathrm{ms}$, when the reaction triggered by the heating pulse has already stopped, the background heating leads to monotonically increasing  mean cloud size.
From our results in fig.~2a we conclude that a temperature variation of 1\% leads to $C_2/ R_2$ variations of at most a few percent, which is negligible with respect to our current measurement accuracy.

In view of these {complex} dynamics we have set up a {system} of coupled differential equations that describe in a more complete fashion the various reaction/loss processes at varying temperatures (see Supplementary Material).
The solid curve in fig. \ref{fig:TDepRates}b) is a result of these calculations which in general show very good agreement with our measurements.

 \begin{figure}[tbp]
	\centering
	\includegraphics[width=1\textwidth]{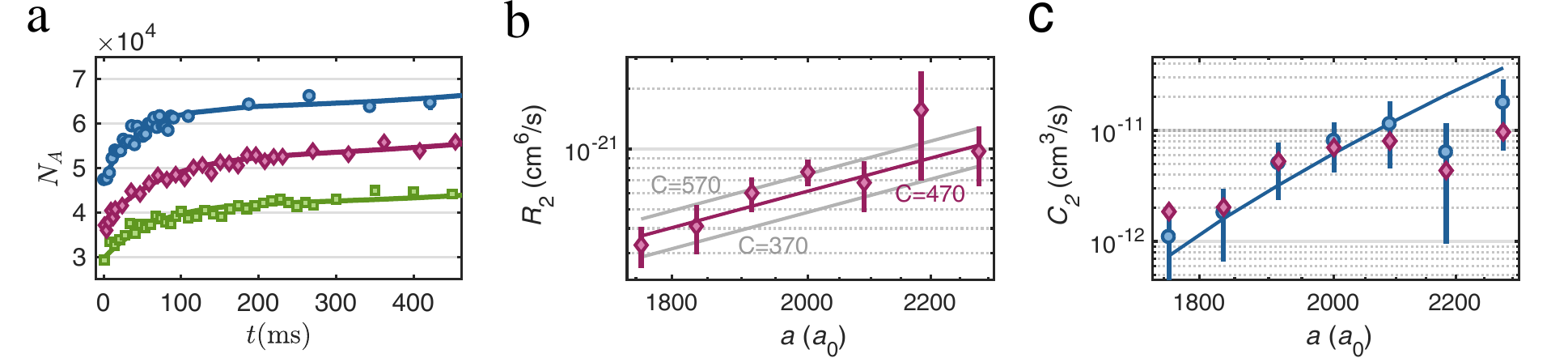}
	\caption{
		\textbf{Dependence of the reaction rate constants on the scattering length. a,} Reaction dynamics for three
		scattering lengths of $a =(1760,1920,2000)a_0$ (green squares, red diamonds and blue circles), corresponding to
		the magnetic fields $B=(705,711,714)\,\mathrm{G}$, respectively.
		The continuous lines are fits based on eq.~\ref{eq:RateCh2I} from which $R_2$ and $C_2$ can be extracted. \textbf{b,} The measured association rate constant $R_2$ as a function of $a$ (red diamonds). $R_2$ roughly follows the universal relation $R_2=C\hbar a^4 / m$, with $C=470$ obtained from a fit to the data (red continuous line). The majority of the data lies in a band between the curves with $C=370$ and $570$ (grey lines).
		\textbf{c,} The dissociation rate constant $C_2$ (red diamonds) as a function of $a$. The temperatures between the individual measurements varied by about $15\%$. To compensate the influence of the temperature we use eq.~\ref{eq:Universal} to rescale $C_2$ to values corresponding to $T=1.5\,\mathrm{\upmu K}$ (blue circles).
		The error bars correspond to temperature uncertainties and the $95\%$ confidence bounds determined by fits as in \textbf{a}. The blue continuous line is the theoretical prediction of eq. \ref{eq:Universal}
 for a  universal constant of $C=470$. }
	\label{fig:RateConstantResults}
\end{figure}

Finally, we investigate the influence of the interaction strength between the particles on the reaction dynamics.
For this, we tune the scattering lengths with the help of the magnetic $B$-field. We note that the dimer-dimer scattering length $a_{dd}$ is given by $a_{dd}=0.6a$, where $a$ is the scattering length for atom-atom collisions \cite{Pet04}.
Figure~\ref{fig:RateConstantResults}a shows three  measurements for $a =(1760,1920,2000)\,a_0$.
For technical reasons, we start with three different $N_A$ at $t=0$.
However, this has negligible influence on the dynamics of the dissociation, which we have checked with a numerical calculation.
Already from the data shown in  fig.~\ref{fig:RateConstantResults}a it is  obvious that the dissociation rates strongly increase with the scattering length.
From fits to these and additional measurements we extract $R_2(a)$ and $C_2(a)$ for various scattering lengths and plot the results on a double logarithmic scale in fig. \ref{fig:RateConstantResults}b, c (red diamonds).
%\replaced[id=dh]{
%We find that $R_2$ increases by a factor of two and almost linearly on this scale in our parameter range. 

The association (three-body recombination) process characterized by the rate constant $R_2$ has been extensively discussed for various Efimovian systems\cite{Fer11,Pir14,Tun14}, where it depends on the three-body parameter, and for non-Efimovian Fermi-Fermi mixtures, where it is suppressed in the low-energy limit\cite{Pet03,DIn05}. By contrast, here we are dealing with a non-Efimovian system of three distinguishable particles, for which a simple dimensional analysis \cite{Fed96} predicts the low-energy threshold law $R_2=C\, \hbar \, a^4/m$, where $C$ is   a universal constant. In fig. \ref{fig:RateConstantResults}b this $a^4$ scaling behavior is plotted for $C = 470$, obtained from a fit to our data. 
Our  results show quite good agreement with the expected power law dependence within the error bars.
 Figure~\ref{fig:RateConstantResults}c shows $C_2$ for various $a$ (red diamonds). These data are % still
% For $C_2(a_{dd})$ we have to consider that the result for $C_2$ is
 still raw in the sense that each measurement is taken at a slightly different temperature which increases with the scattering length (see Supplementary Material for details). 
In order to compensate this temperature change we use eq.~\ref{eq:ClTDep} to rescale the measured $C_2$ rate constants to values corresponding to a constant temperature $T=1.5\,\mathrm{\upmu K}$ (see blue circles in fig. \ref{fig:RateConstantResults}b).  The resulting rate constant $C_2$ increases by more than one order of magnitude in the tuning range and agrees reasonably with the  theoretical prediction (without any free parameter), 
\begin{equation}
C_2=C\,\,\frac{\left(\pi m k_B\right)^{1/2}k_B}{2h^{2}}\,\,\,T^{3/2}\,\, \,a^4 \, e^{-\frac{E_b(a)}{k_B T}},
\label{eq:Universal}
\end{equation}
which is obtained by inserting $R_2=C\, \hbar \, a^4/m$  into  eq.~\ref{eq:ClTDep} and using again $C = 470$.
As far as we know there is no direct theoretical prediction for this number.
D'Incao and co-workers \cite{DIn09} calculated dimer-dimer elastic and inelastic
scattering properties in a wide range of collision energies. For the energy
interval relevant here, these calculations indicate $30\lesssim C\lesssim
100$ which is also consistent with our own numerical estimates based on
\cite{Pet04,Pet05}.
The large  discrepancy between the theoretical and experimental value needs to be investigated in future studies.
%\replaced[id=dh2]{result of}{}\replaced[id=dh]{$C=470$,}{} \replaced[id=dh2]{and will be target of}{which will be investigated in detail in} \replaced[id=dh]{future studies.}{
%Finally we note that we have  performed ab-initio, four-body calculations for the universal constant $C$, which currently result in $C\approx25$. Thus there is a discrepancy with respect to the experimentally determined value of 
% $C\approx530$, which might partially be explained by the fact that currently  
% $p$-wave attraction in the atom-dimer collision is not yet included in the theory. This will be investigated in detail in future. }

In conclusion, we have investigated the collisional dissociation of ultracold molecules in a single reaction channel
which is characterized by the precisely defined quantum states of the involved atoms and molecules.
Using a heating pulse we shift an atom/ molecule mixture which is initially in detailed balance out of equilibrium and measure the evolution of the system  until it reaches a new equilibrium.
This allows us to determine reaction rate constants, in particular for the collisional dissociation of two molecules.
Furthermore, we find a strong temperature dependence of this rate which is consistent with the well known Arrhenius equation.
In addition, we find  agreement of the association (dissociation) rate constant with a scaling behavior of  $a^4$ ($a^4\,e^{-E_b/k_BT}$), respectively.
%{for the dissociation reaction rate constant with the $a_{dd}^4\,e^{-E_b/k_BT}${} scaling behaviour, where $a_{dd}$ is the scattering length for collisions between two molecules.}
From our data we estimate the universal constant $C\approx470$, which 
is in discrepancy with the theoretical prediction.
%ects the association/dissociation rate constants to the scattering length. \replaced[]{However, the constant is in large discrepancy with $C=25$ obtained from ..... In future work, this discrepancy .
% \replaced[id=dh2]{This discrepancy will be studied in detail in future work.}{We plan to investigate this discrepancy in the near future.}
For the future, we plan to extend the current work to study the dynamics of chemical reactions in a regime, where Fermi and Bose statistics play an important role.

\newpage
\section*{Methods}
\subsection*{Preparation of the atomic and  molecular quantum gas}
To prepare our sample of ultracold atoms and molecules, we initially trap $10^9$ $^6$Li atoms in a magneto-optical trap, where the atoms are cooled to a temperature of $700\,\upmu \mathrm{K}$.
The particles are transferred to an optical dipole trap of a focused 1070nm laser beam with an efficiency of $1\%$.
To generate a balanced distribution ($50\%/50\%$) of atoms in the $|m_F=\pm1/2>$ spin states we apply a resonant $100\,\mathrm{ms}$ radio frequency pulse.
Initially the optical trap has a depth of $4\,\mathrm{mK}\times k_B$  and %\replaced[id=dh]
is subsequently ramped down within $6\,\mathrm{s}$ to $1.3\,\mathrm{\upmu K}\times k_B$  to perform forced evaporative cooling.
This is carried out at a magnetic field of $780\,$G and during this process Feshbach molecules  form via three-body recombination.
To suppress particle loss in the experiments and to assure harmonicity of the trapping potential, the trap depth is ramped up again to $U_{0}= 21\,\mathrm{\upmu K}\times k_B$ after evaporation.
We then ramp the $B$-field in a  linear and adiabatic fashion to the specific value at which the experiment will be carried out, within the range of $705\,\mathrm{G}$ to $723\,\mathrm{G}$.
After a holding time of $100\,\mathrm{ms}$ the gas has a temperature of approximately $1.2$ to $1.3\,\mathrm{\upmu K}$ and  is in chemical equilibrium, with 80\% to 90\% of all atoms being bound in Feshbach molecules.
The binding energy of the molecules can be determined from\cite{Jul14}  $E_b=\frac{\hbar^2}{m(a-\bar{a})^2}\left(1+2.92\frac{\bar{a}}{a-\bar{a}}-0.95\frac{\bar{a}^2}{(a-\bar{a})^2}\right)$ using $\bar{a}=29.9a_0$, which yields values between $6$ and $10\,\mathrm{\upmu K}\times k_B$ in our B-field range. The scattering length $a$ as a function of the B-field is taken from ref.\cite{Zur13}. It can be approximated with $a= a_{bg}\left(1-\frac{\Delta B}{B-B_0}\right)$, where $\Delta B= -263.3\,\mathrm{G}$ is the width of the resonance and $a_{bg}=-1582\,a_0$ is the background scattering length.

\subsection*{Removing of Feshbach molecules}
To optically pump the Feshbach molecules into undetected states, we use a {673nm} laser
with a peak intensity of $I_0=500\mathrm{mW/cm^2}$ which excites all Feshbach molecules to the $A^1\Sigma_u^+,v'=68$ state \cite{Pat05} (see fig. 1c) within  $500\mathrm{\mu s}$.
 % via the $A^1\Sigma_u^+,v'=68$ state (see fig. 1c) we use a pulsed Gaussian laser beam with peak intensity $I_0=500\mathrm{mW/cm^2}$, a beam waist $w=1.1\mathrm{mm}$ and a pulse length of $t=500\mathrm{\mu s}$.
%The Feshbach molecules are fully removed from the sample within this pulse duration.
The excited molecular state  decays within a few ns either into two unbound atoms which quickly leave the trap or into deeply bound Li$_2$ molecules which are invisible for our detection.

\subsection*{Acknowledgements}
The authors thank C. Chin for fruitful discussions, W. Schoch and B. Deissler for their work in the early stages of the experiment, and J. D'Incao for providing us with the data of ref.\cite{DIn09}
This work was supported by the German research foundation Deutsche Forschungsgemeinschaft (DFG) within SFB/TRR21 and by
the Baden-W\"{u}rttemberg Foundation and the Center for Integrated Quantum Science and Technology (IQST), and by the European Research Council (FR7/2007-2013 Grant Agreement No. 341197).

\clearpage

\section*{Supplementary Information}
\renewcommand\thefigure{S\thesection\arabic{figure}}
\renewcommand\thetable{S\thesection\arabic{table}}

\newcolumntype{C}[1]{>{\centering\arraybackslash}p{#1}}

\setcounter{figure}{0}
\setcounter{equation}{0}

\subsection*{Modeling of the reaction dynamics }
We perform model calculations to describe in more detail the measured dynamics of the atom/ molecule system, triggered by the initial heat pulse.
% to get the time evolution triggered by the parametric excitation.
For this, we integrate  the following  coupled system of rate equations for the atom number $N_A$, the molecule number $N_M$ and the temperature $T$,

\begin{equation}
\dot{N}_A=C_2 \, a_1\frac{N_M^2}{\sigma_r^2\, \sigma_{ax}}-R_2 \, a_2\frac{N_M \, N_A^2}{\sigma_r^4\sigma_{ax}^2}
\label{eqsup:SysDiff1}
\end{equation}
\begin{equation}
\dot{N}_M=-\frac{\dot{N}_A}{2}-C_{DD} \, a_1\frac{N_M^2}{2 \, \sigma_r^2 \, \sigma_{ax}}
\label{eqsup:SysDiff2}
\end{equation}

\begin{equation}
\dot{T}= -\frac{E_b}{6k_B \, (N_A+N_M)}\dot{N}_A+C_H
\label{eqsup:SysDiff3}
\end{equation}

where  $a_1=(4\pi^{3/2})^{-1}$ and $a_2=2^{-7}(2\pi^{2})^{-3/2}$ are numerical constants.
Equation (\ref{eqsup:SysDiff1}) is identical to eq.~(2) in the main text.
The first term in eq.~(\ref{eqsup:SysDiff2}) corresponds to the conversion between molecules and unbound atoms, while the second term accounts for molecule losses in inelastic dimer-dimer collisions with rate constant of $C_{DD}=2.3\times10^{-13}\,\mathrm{cm^3/s}$, which is extracted from the previous measurement of ref. \cite{SBou04}.
Equation~(\ref{eqsup:SysDiff3}) has two contributions.
The first one accounts for cooling due to endothermic dissociation and heating due to exothermic recombination reactions.
The second contribution corresponds to background heating of the gas caused, e.g., by off-resonant scattering of the dipole-trap light.
Equations (\ref{eqsup:SysDiff1}) and  (\ref{eqsup:SysDiff2}) are coupled via the cloud sizes $\sigma_{r(ax)}=\sqrt{k_BT/m_M\omega_{r(ax)}^2}$ to the temperature equation (\ref{eqsup:SysDiff3}).

\begin{figure}[tbp]
 	\centering
 	\includegraphics[width=.7\textwidth]{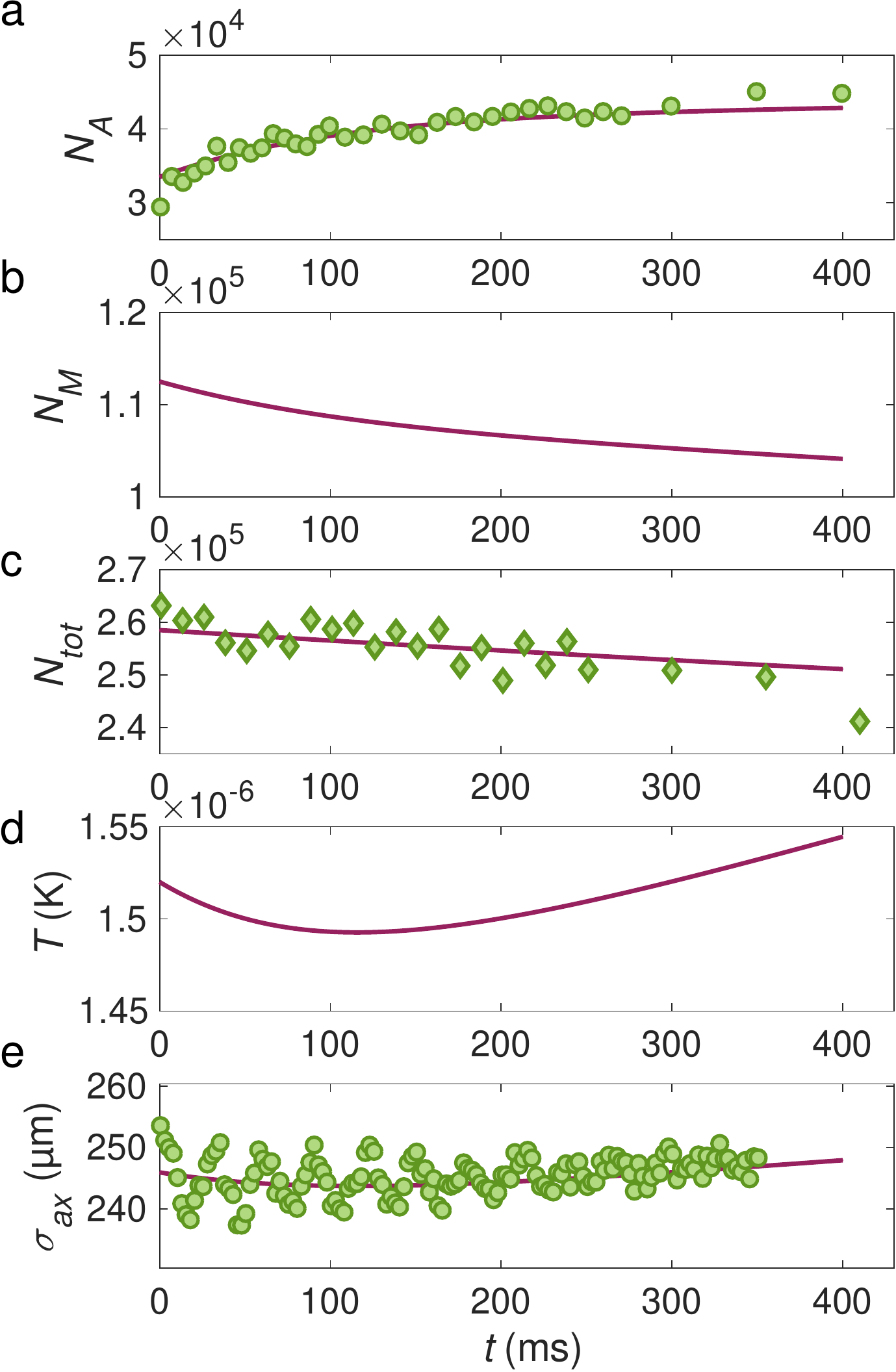}
 	\caption{\textbf{Model calculation for the evolution of the atom/ molecule system.}
		Results from the coupled differential equations  (\ref{eqsup:SysDiff1}-\ref{eqsup:SysDiff3}) (continuous lines).
Plot symbols show experimental data for $709\,\mathrm{G}$.  For details see text. }
 	\label{fig:CoupledSys}
 \end{figure}

The results of a corresponding calculation at $709\,\mathrm{G}$ are shown in fig. \ref{fig:CoupledSys} and are in very good agreement with the experimental data.
The values of the parameters $C_2$ and $R_2$ are the same as in the main text. $C_H=(3.0\pm1.0)\times10^{-7}\,\mathrm{s^{-1}}$ is mainly determined by the long-time evolution of the cloud size  (see fig. \ref{fig:CoupledSys}d,e).

The total particle number in  fig.~\ref{fig:CoupledSys}c) exhibits the losses caused by the inelastic dimer-dimer collisions and agrees well with the
experimental measurements. % at $709\,\mathrm{G}$ given by the circles.
The temperature in fig. \ref{fig:CoupledSys}d) first decreases slightly due to endothermic dissociation after the heat pulse and then increases again
due to the dipole-laser induced photodissociation.
Figure \ref{fig:CoupledSys}e  shows the calculated cloud width $\sigma_{ax}$ as determined by the temperature $T$. It agrees well with the measurements (green circles) if we average over the small-amplitude collective oscillations which have been excited by the initial heat pulse.

\subsection*{Thermometry}

   \begin{figure}[tbp]
	\centering
	\includegraphics[width=.8\textwidth]{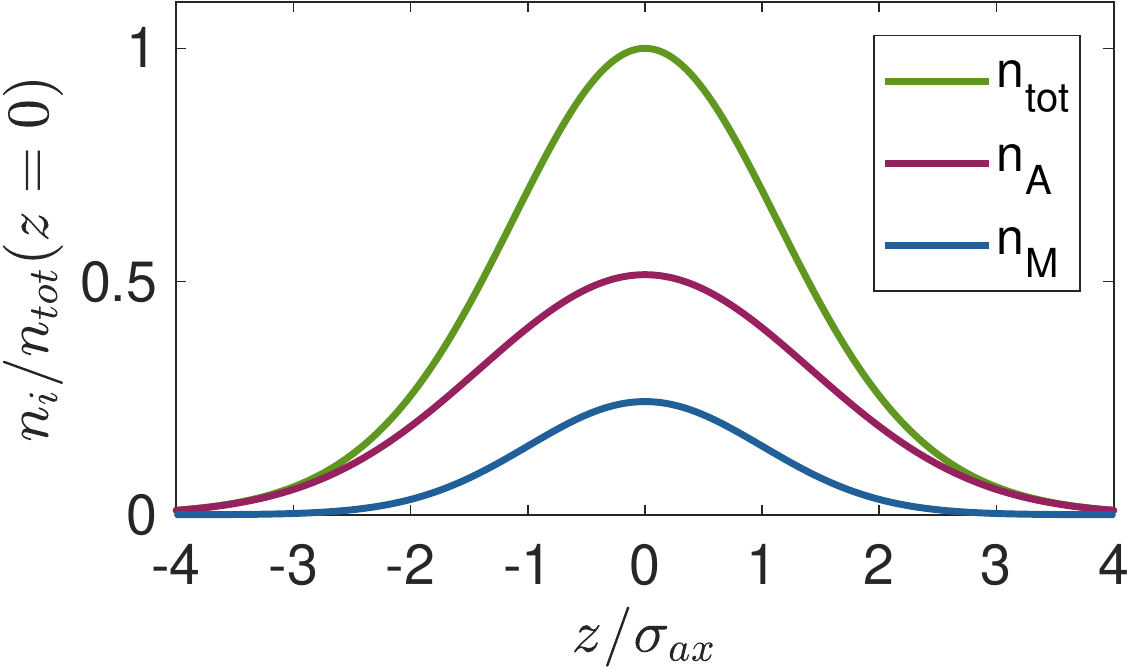}
	\caption{\textbf{Calculated 1D density distribution for a thermal cloud of atoms and molecules.}
In a harmonic potential non-interacting atoms and molecules exhibit a Gaussian density distributions $n_A$ and $n_M$, respectively, of which the widths differ by a factor or $\sqrt{2}$. The total density distribution is the sum $n_{tot}=2n_M+n_A$. Here, the atom fraction $N_A/N_{tot}$ is $0.6$.}
	\label{supfig:Thermo}
\end{figure}

Within the parameter range of our experiments the  molecular and atomic density distributions are each well described by that of a non-interacting thermal gas located in a harmonic trap. The axial sizes of the  molecular and atomic clouds are given by $\sigma_{ax}=\sqrt{k_BT/m\omega_{ax}^2}$ and
$\sigma_{ax,A} = \sqrt{2}\sigma_{ax}$, respectively. Because the axial trapping frequency $\omega_{ax}$ is precisely known for our setup, we can determine the temperature $T$ by measuring the molecular or atomic cloud size. These cloud sizes are extracted from images of the mixed atom/ molecule clouds after careful analysis, for which we also determine the atom fraction $N_A/N_{tot}$ of the cloud. Figure~\ref{supfig:Thermo} shows a calculated typical example for the 1D density distributions (where the transverse directions have been integrated out) for atoms $n_A$, molecules $n_M$ and both $n_{tot}=2n_M+n_A$.

\subsection*{Rescaling the measured $C_2$ to a constant temperature}
Table \ref{tab:Temps} is the list of temperatures at which the $C_2$ measurements in fig.~\ref{fig:RateConstantResults}c
are taken. The statistical uncertainty of the temperatures is around $\Delta T =  0.08\,\mathrm{\upmu K}$.
%\replaced[]{In fig. \ref{fig:RateConstantResults} we show rescaled results for the dissociation rate constant.
%To rescale the data we use the temperature dependence of eq. \ref{eq:Universal} and the temperatures measured directly after the excitation pulse.
%The temperatures are given in following table.}{}
 \begin{table}[h!]
 	\caption{Temperatures and scattering lengths of the measurements in fig. \ref{fig:RateConstantResults}c.}	
 	\begin{tabular}{l r r r r r r r r}
\toprule
 		\multicolumn{1}{ c }{$a$ ($a_0$)} \, \, \,  &
 		\multicolumn{1}{ c }{$1760$} &
 		\multicolumn{1}{ c }{$1830$} &
 		\multicolumn{1}{ c }{$1920$} &
 		\multicolumn{1}{ c }{$2000$} &
 		\multicolumn{1}{ c }{$2090$} &
 		\multicolumn{1}{ c }{$2180$} &
 		\multicolumn{1}{ c }{$2280$} &\\
 		$T$ ($\upmu \mathrm{K}$)&$1.60$ &$1.52$  &$1.51$ &$1.47$  &$1.42$  &$1.41$  &$1.35$ \\
 		\hline
 	\end{tabular}
 	\label{tab:Temps}
 \end{table}
%\replaced[id=dh]{
The temperatures increase with decreasing scattering length $a$. 
This temperature change is a result of the way we prepare the sample. In particular,  the  magnetic field ramp to the target field takes place
within a $21\,\mathrm{\upmu K}\times k_B$ deep trap which prevents further evaporative cooling.
With decreasing $a$, the binding energy of the dimers increases and therefore for a given temperature the equilibrium molecule fraction increases. The corresponding molecular association, however, heats the sample. 

%The smaller $a_{dd}$ at the target field, the larger the number of molecules produced and therefore release energy deposited in the gas.}{}
%\replaced[id=dh]{To compensate the influence of the differing temperatures, we rescale $C_2$ to a value that corresponds to a measurement at $T=1.5\,\mathrm{\upmu K}$. }$.}{}

\end{document}